\documentstyle[amssymb,preprint,aps]{revtex}\tightenlines
\begin{document}
\draft
\title{The response function of a Hall magnetosensor}
\author{Y. G. Cornelissens and F.M. Peeters$%
%TCIMACRO{\UNICODE[m]{0xb0}}%
%BeginExpansion
{{}^\circ}%
%EndExpansion
$}
\address{Departement Natuurkunde, Universiteit Antwerpen (UIA),\\
Universiteitsplein 1, B-2610 Antwerpen, Belgium}
\date{\today}
\maketitle

\begin{abstract}
Patterned two-dimensional electron gas (2DEG) systems into micrometer Hall
bars can be used as Hall magnetosensors to provide detailed information on
the magnetic field distribution. In this way, ballistic Hall probes have
already been studied and used successfully. Here, the response function of a
Hall sensor is determined in the diffusive regime, which allows this device
to be used as a magnetosensor for the determination of inhomogeneous
magnetic field distributions. Furthermore, the influence of the geometry of
the Hall bar on this response function, such as circular corners and
asymmetry in the probes, is also investigated and appears to be
non-negligible.
\end{abstract}

\section{Introduction}

The Hall effect, discovered in 1879 \cite{Hall}, was successfully used to
study two-dimensional electron gas (2DEG) systems, which resulted in two
Nobel Prizes (in 1985 and 1998, respectively) for the observation of the
Integer \cite{integer} and the Fractional \cite{frac} Quantum Hall Effect.
On the other hand, the Hall effect can also be used to provide detailed
information on the magnetic field distribution, allowing these 2DEG systems
to be used as Hall magnetosensors or Hall sensors \cite{sensors}. The recent
micrometer Hall sensors have considerable advantages compared to other
magnetic field measurements, like its noninvasive character, its high
magnetic field sensitivity, the small dimensions of the active region and
the broad range of temperature and magnetic field strength within which it
can be used. Hall probes have previously been used in experiments to study
magnetic flux profiles using scanning Hall probe microscopy \cite{SHPM}, and
were successfully applied for time and space resolved detection of
individual vortices in superconductors \cite{PRB46,PRL71}. Furthermore, they
are becoming increasingly popular as an alternative for memory devices
(MRAM) \cite{mram,hybrid,monzon,jonas1,jonas2}. Recently, submicron
ballistic Hall probes were successfully used to investigate the
thermodynamic properties of submicron superconducting and ferromagnetic\
disks \cite{geim}.

\noindent In order to improve the resolution of these Hall magnetometers, it
is necessary to provide a quantitative theory which relates the experimental
data, in terms of resistance and voltage measurements, to the properties of
the magnetic field, and more precisely to the size and strength of the
inhomogeneities in this magnetic field. In other words, it is necessary to
determine the {\it Hall response function}. The theory for the Hall
magnetometer in the ballistic regime was given in Ref. \cite{ballistic},
where it was found that for small magnetic field strength the Hall
resistance is determined by the average magnetic field in the Hall junction (%
$R_{H}=\alpha ^{\ast }\left\langle B\right\rangle $, with $\alpha ^{\ast }$
the effective Hall coefficient), and that it is rather insensitive to the
exact position of the magnetic field inhomogeneity. This is in contrast to
the diffusive regime, where scattering processes strongly determine the
electron transport. In the latter case there is no longer a simple relation
between the Hall resistance and the magnetic field inhomogeneity \cite
{diffusive}, and the Hall resistance depends much more sensitively on the
above mentioned factors. Previous studies \cite
{ballistic,diffusive,diffusive2} found that the geometry of the device has
considerable influence on the Hall response.

The aim of the present paper is to determine the response function of a Hall
sensor in the diffusive regime (e.g. this is the regime at room
temperature), for given dimensions and geometric characteristics of the
device.

This paper is organised as follows. In Sec. II we describe the numerical
approach used to determine the Hall response in the diffusive regime, and in
Sec. III we give the numerical results for a symmetric Hall cross. The
influence of asymmetric geometries of the Hall bar is investigated in Sec.
IV, and our conclusions are presented in Sec. V.

\section{Numerical approach}

In order to describe the transport properties of the 2DEG in the diffusive
regime we start from the continuity equation

\begin{equation}
\overrightarrow{\nabla }\cdot \overrightarrow{J}+\frac{\partial \rho }{%
\partial t}=0\text{,}  \eqnum{1}  \label{eq1}
\end{equation}
with $\overrightarrow{J}$ and $\rho $ the current and charge density,
respectively. In the steady state we have $\partial \rho /\partial t=0$ and $%
\overrightarrow{\nabla }\times \overrightarrow{E}=0$, which implies that $%
\overrightarrow{E}=-\overrightarrow{\nabla }\phi $, with $\phi $ the
potential. For linear transport we use Ohm's law $\overrightarrow{J}=\sigma 
\overrightarrow{E}$, where $\sigma $ is the conductivity tensor, which
reduces Eq.~(\ref{eq1}) to the following two-dimensional partial
differential equation $\overrightarrow{\nabla }\cdot \left[ \sigma \left(
x,y\right) \overrightarrow{\nabla }\phi \left( x,y\right) \right] =0$, which
can be written more explicitly as

\begin{equation}
\frac{\partial }{\partial x}\left( \sigma _{xx}\frac{\partial \phi }{%
\partial x}+\sigma _{xy}\frac{\partial \phi }{\partial y}\right) +\frac{%
\partial }{\partial y}\left( \sigma _{yy}\frac{\partial \phi }{\partial y}%
+\sigma _{yx}\frac{\partial \phi }{\partial x}\right) =0\text{,}  \eqnum{2}
\label{eq2}
\end{equation}

\noindent with $\sigma _{xx}=\sigma _{yy}=\sigma _{0}/\left( 1+\left[ \mu
B\left( \overrightarrow{r}\right) \right] ^{2}\right) $ and $\sigma
_{xy}=-\sigma _{yx}=\mu B\left( \overrightarrow{r}\right) \sigma _{xx}$,
where $\sigma _{0}=n_{e}e\mu $ is the zero field conductivity, with $\mu $
the mobility and $n_{e}$ the electron density of the 2DEG. Eq.~(\ref{eq2})
is then solved numerically for given boundary conditions using the finite
difference method. The boundary conditions reflect the geometry of the Hall
bar (Fig.~\ref{Fig1}), and will vary as different geometries are
investigated. The numerical approach presented here is more general than the
one presented in Ref. \cite{diffusive2}, where the effect of probe geometry
on the Hall response in the limit of very weak inhomogeneous magnetic fields
was studied. Taking the limit of small magnetic fields in Eq.~(\ref{eq2}),
i.e. $\sigma _{xx}=\sigma _{yy}=\sigma _{0}$ and $\sigma _{xy}=-\sigma
_{yx}=\mu B\left( \overrightarrow{r}\right) \sigma _{0}$, leads in fact to
Eq. (5) of Ref. \cite{diffusive2} with $\sigma _{0}=\sigma _{B}$. We found
that the use of such a small magnetic field expansion leads to an
oversimplification and misses some essential physics of the device.

When a spatially inhomogeneous magnetic field $B(x,y)$ is present the Hall
resistance will in general depend on the exact position of the inhomogeneous
field distribution with respect to the Hall cross. This response of the Hall
cross is described by a {\it response function} $F_{H}(x,y)$ from which we
obtain the Hall resistance

\begin{equation}
R_{H}=-\frac{1}{n_{e}e}\frac{\int dx\int dy\;F_{H}\left( x,y\right)
\;B\left( x,y\right) .}{\int dx\int dy\;F_{H}\left( x,y\right) }  \eqnum{3}
\label{eq3}
\end{equation}

\noindent In the limiting case of a delta-shaped magnetic field profile: $%
B\left( x,y\right) =\Phi \delta \left( x-x_{0}\right) \delta \left(
y-y_{0}\right) $, with $\Phi =B_{0}S$ the flux through an area $S$, Eq.~(\ref
{eq3}) leads to the following response: 
\begin{equation}
R_{H}=\left( -\Phi /n_{e}e\right) \;\frac{F_{H}\left( x_{0},y_{0}\right) }{%
\int dx\int dy\;F_{H}\left( x,y\right) }.  \eqnum{4}
\end{equation}
Consequently, 
\begin{equation}
\widetilde{F}_{H}(x_{0},y_{0})=\frac{F_{H}\left( x_{0},y_{0}\right) }{\int
dx\int dy\;F_{H}\left( x,y\right) }=\frac{R_{H}}{\left( -\Phi /n_{e}e\right) 
}  \eqnum{5}
\end{equation}
is the spatially dependent normalized Hall response of a delta-function
magnetic field profile, expressed in units of area$^{-1}$. In this way it is
possible to determine the response function of the Hall cross by placing
this delta-function magnetic profile in every point of the cross and
calculating the Hall resistance. In practice, a delta function magnetic
field distribution can not be used, as it is impossible to generate
experimentally, and furthermore, because we have to solve Eq.~(\ref{eq2})
numerically. Therefore, we shall approximate this profile with a magnetic
step of radius $r_{0}$ and strength $B_{0}$: $B\left( x,y\right)
=B_{0}\theta \left( \left| \overrightarrow{r}-\overrightarrow{r}^{\prime
}\right| -r_{0}\right) $, which is also called a {\it magnetic dot}, and
which approximates a flux tube of radius $r_{0}$ and flux $\Phi =B_{0}\pi
r_{0}^{2}$, centered around position $\overrightarrow{r}^{\prime }$.

Once the response function is known, one can scan the sample with the Hall
sensor and measure the Hall resistance $R_{H}\left( x_{i},y_{j}\right) $ in
every point $\left( x_{i},y_{j}\right) $ of the sample. The magnetic field
distribution $B(x,y)$ is then obtained by performing an inverse
transformation. Eq.~(\ref{eq3}) can be written as follows: 
\begin{equation}
R_{H}(x_{i},y_{j})=-\frac{1}{n_{e}e}\int dx\int dy\;\widetilde{F}%
_{H}(x,y)\;B\left( x_{i}+x,y_{j}+y\right) \text{,}  \eqnum{6}
\end{equation}
with $(x_{i},y_{j})$ the position of the Hall probe on the surface to be
scanned (Fig.~\ref{Fig2}). In polar coordinates this equation becomes 
\begin{equation}
R_{H}(x_{i},y_{j})=R_{H}\left( r_{ij},\theta _{ij}\right) =-\frac{1}{n_{e}e}%
\int dr\int d\theta \;\widetilde{F}_{H}\left( r,\theta \right) \;B\left(
r^{\prime },\theta ^{\prime }\right) \text{,}  \eqnum{7}  \label{eq7}
\end{equation}
with $r^{\prime }=\sqrt{r^{2}+r_{ij}^{2}+2rr_{ij}\cos \left( \theta
_{ij}-\theta \right) }$, and $\theta ^{\prime }=\arctan \left( \frac{r\sin
\theta +r_{ij}\sin \theta _{ij}}{r\cos \theta +r_{ij}\cos \theta _{ij}}%
\right) $. Using vector notation we can write the Fourier transform: 
\begin{eqnarray}
R_{H}(\overrightarrow{k}) &=&\int d\overrightarrow{r}_{ij}\exp \left( -i%
\overrightarrow{k}.\overrightarrow{r}_{ij}\right) R_{H}\left( 
\overrightarrow{r}_{ij}\right)  \eqnum{8}  \label{eq8} \\
&=&\int d\overrightarrow{r}_{ij}\exp \left( -i\overrightarrow{k}.%
\overrightarrow{r}_{ij}\right) \left\{ \int d\overrightarrow{r}\;\widetilde{F%
}_{H}\left( \overrightarrow{r}\right) \;B\left( \overrightarrow{r}+%
\overrightarrow{r}_{ij}\right) \right\}  \nonumber \\
&=&\int d\overrightarrow{r}\;\exp \left( +i\overrightarrow{k}.%
\overrightarrow{r}\right) \widetilde{F}_{H}\left( \overrightarrow{r}\right)
\int d\left( \overrightarrow{r}+\overrightarrow{r}_{ij}\right) \exp \left( -i%
\overrightarrow{k}.\left( \overrightarrow{r}+\overrightarrow{r}_{ij}\right)
\right) B\left( \overrightarrow{r}+\overrightarrow{r}_{ij}\right)  \nonumber
\\
&=&\widetilde{F}_{H}\left( -\overrightarrow{k}\right) B\left( 
\overrightarrow{k}\right)  \nonumber
\end{eqnarray}
And consequently 
\begin{equation}
B\left( \overrightarrow{k}\right) =\frac{R_{H}(\overrightarrow{k})}{%
\widetilde{F}_{H}\left( -\overrightarrow{k}\right) }\text{.}  \eqnum{9}
\end{equation}

\noindent The magnetic field distribution is then obtained by performing the
inverse Fourier transform: 
\begin{equation}
B\left( \overrightarrow{r_{ij}}\right) =\frac{1}{2\pi }\int d\overrightarrow{%
k}\exp \left( i\overrightarrow{k}.\overrightarrow{r_{ij}}\right) B\left( 
\overrightarrow{k}\right) =\frac{1}{2\pi }\int d\overrightarrow{k}\exp
\left( i\overrightarrow{k}.\overrightarrow{r}_{ij}\right) \frac{R_{H}(%
\overrightarrow{k})}{\widetilde{F}_{H}\left( -\overrightarrow{k}\right) }%
\text{.}  \eqnum{10}  \label{eq10}
\end{equation}

\section{Symmetric Hall cross}

The system we envisage is given schematically in Fig.~\ref{Fig1}, a Hall bar
with four identical leads. A voltage drop $V_{0}$ across the Hall bar along
the $y$ direction\ generates a current $I$. The presence of the magnetic dot
will then give rise to a potential profile such as the one shown in Fig.~\ref
{Fig3}, from which the longitudinal resistance $R_{L}=V_{0}/I$ and the Hall
resistance $R_{H}=V_{H}/I$ can be calculated. In order to allow immediate
comparison with any given experimental setup we scale our physical
quantities as follows: lengths in units of the probe width $W$, magnetic
field strength in units of the inverse mobility of the 2DEG, $\mu ^{-1}$,
voltages in units of $V_{0}$, and resistances in units of the zero magnetic
field resistivity $\rho _{0}=n_{e}e\mu $. The normalized Hall response
function $\widetilde{F}_{H}(x,y)$ is then given in units of $1/W^{2}$, i.e.
the inverse of the Hall junction area.

The response function 
\begin{equation}
\widetilde{F}_{H}(x_{i},y_{j})=\frac{F_{H}\left( x_{i},y_{j}\right) }{\int
dx\int dy\;F_{H}\left( x,y\right) }=R_{H}/\left( B_{0}\pi r_{0}^{2}\right) 
\eqnum{11}  \label{eq11}
\end{equation}
is numerically calculated for a magnetic dot placed in every grid point $%
\left( x_{i},y_{j}\right) $ in the Hall bar. As an example we took $\mu
B_{0}=0.2$ and $r_{0}^{\prime }=r_{0}/W=0.1$. When moving the dot along the
middle of the voltage and current probes we obtained a nearly identical
response, given in Fig.~\ref{Fig4} for both cases ($\theta =0^{\text{o}}$:
solid dots; $\theta =90^{\text{o}}$: open dots). Both responses can be very
closely represented by the following relation in polar coordinates: 
\begin{equation}
\widetilde{F}_{H}\left( r,\theta \right) =A\;\frac{1}{1+\left( Cr\right) ^{4}%
}\text{,}  \eqnum{12}  \label{eq12}
\end{equation}
with $(A,C)=(0.4896,1.267)$ for $\widetilde{F}_{H}\left( r,\theta =0^{\text{o%
}}\right) $ (dotted curve in Fig.~\ref{Fig4}), and $(A,C)=(0.4896,1.303)$
for $\widetilde{F}_{H}\left( r,\theta =90^{\text{o}}\right) $ (dashed curve
in Fig.~\ref{Fig4}).\ Along $(r,\theta =0^{\text{o}})$ and $(r,\theta =90^{%
\text{o}})$ the response is, within 2\%, constant over the range $r<0.2W$.
Near the edge of the Hall cross and inside the probes we find a very rapid
decrease of $\widetilde{F}_{H}$.

However, when calculating the Hall response for a magnetic dot placed in
every grid point of the Hall cross it is clear from the contourplot in Fig.~%
\ref{Fig5} that this rather simple relation~(\ref{eq12}), can not be upheld
when the angle $\theta $ of the cross-section is different from $0^{\text{o}%
} $ or $90^{\text{o}}$. Notice that the response function is slightly skewed
which was not reproduced by the linear (in magnetic field) theory of Ref.~ 
\cite{diffusive2}. As can be seen from Fig.~\ref{Fig5}, the Hall response
function exhibits inverse symmetry with respect to the center of the cross,
but is not axially symmetric, as for $\theta =135^{\text{o}}$ two maxima
appear at opposite corners of the junction at $r/W\approx 0.5$. We found
that this asymmetry is also present for small magnetic fields, and
disappears only for extremely low fields, i.e. $B_{0}\rightarrow 0$. The
broken $x-y$ symmetry arises due to the choice of specific probes which are
used for injection of the current, in combination with the Lorentz force.

Next, considering only the central circular part of the junction ($r/W\leq
0.5$), the dot is moved along different angles $\theta $. Choosing at first
only the `interesting' directions such as $\theta =0^{\text{o}},45^{\text{o}%
},90^{\text{o}}$ and $135^{\text{o}}$, indicated in Fig.~\ref{Fig5}, we
obtained the response given in Fig.~\ref{Fig6}. These curves could be fitted
over the range $0\leq r/W\leq 0.5$ to the expression 
\begin{equation}
\widetilde{F}_{H}\left( r,\theta \right) =A\;\frac{1+\left( D\left( \theta
\right) r\right) ^{2}}{1+\left( C_{2}\left( \theta \right) r\right)
^{2}+\left( C_{4}\left( \theta \right) r\right) ^{4}}\text{,}  \eqnum{13}
\label{eq13}
\end{equation}
with $D\left( \theta \right) $, $C_{2}\left( \theta \right) $ and $%
C_{4}\left( \theta \right) $ parameters which are a function of the angle $%
\theta $. The values of these parameters are given in Fig.~\ref{Fig7}, as a
function of the angle $\theta $, along with their calculated values for
other angles in the range of $\theta =\left[ 0^{\text{o}},180^{\text{o}}%
\right] $ (solid symbols). Notice that parameter $C_{2}=1$ for $0^{\text{o}%
}<\theta <90^{\text{o}}$ and approaches zero, i.e. $C_{2}=0$ in the range $%
90^{\text{o}}<\theta <180^{\text{o}}$ which is explained by the curvature of
the Hall response for these angle values. This difference in curvature can
also be seen from the values for $D(\theta )$ which also exhibit a
discontinuity for $\theta >90^{\text{o}}$. Furthermore, parameter $C_{4}$
exhibits two distinct minima for $\theta =45^{\text{o}}$ and $\theta =135^{%
\text{o}}$, the two diagonal directions in the Hall junction. Knowing all
these parameter values $\left( A,D,C_{2},C_{4}\right) $ for an arbitrary
direction $\theta $ allows us to describe the Hall response in the central
part of the junction with a single expression (Eq.~(\ref{eq13})) which,
after Fourier transformation, can be inserted in Eq.~(\ref{eq10}) to
calculate the magnetic field distribution.

The parameter $A$ is independent of the direction $\theta $, but is a
function of the strength and size of the magnetic dot which we investigated
simply by placing a magnetic dot with variable strength and size in the
center of the Hall junction. The dependence of the Hall response, in the
center of the Hall cross, on the strength $\mu B_{0}$ and the radius $%
r_{0}^{\prime }$ of the magnetic dot is shown in Fig.~\ref{Fig8} (solid
dots). The numerical results are closely described by the following relation
(solid curves in Fig.~\ref{Fig8}):

\begin{equation}
A=\widetilde{F}_{H}\left( 0,0^{\text{o}}\right) =\frac{A_{1}\left(
r_{0}^{\prime }\right) }{1+\left( \mu B_{0}\ast A_{2}\left( r_{0}^{\prime
}\right) \right) ^{2}}\text{,}  \eqnum{14}  \label{eq14}
\end{equation}
where the dependence of the parameters $A_{1}$ and $A_{2}$ on the radius $%
r_{0}^{\prime }$ is shown in the inset of Fig.~\ref{Fig8} (solid and open
dots, respectively). In this figure we see that $A_{1}=0.493$ remains
constant up to $r_{0}^{\prime }\approx 0.3$ (dotted line), while the
parameter $A_{2}$ can be represented by a linear equation: $%
A_{2}=0.522-0.229\;r_{0}^{\prime }$ (dashed line), likewise up to $%
r_{0}^{\prime }\approx 0.3$. Notice that for $\mu B_{0}<0.5$ the response in
the center of the Hall cross ($\widetilde{F}_{H}\approx 0.49$) does
practically not depend on the magnetic field strength, nor on the radius of
the magnetic dot for $r_{0}^{\prime }\leq 0.4$. Notice also that at $\mu
B_{0}\approx 0.5$ the curves cross, and consequently that, with increasing
radius $r_{0}^{\prime }$, the Hall response decreases for $\mu B_{0}<0.5$
and increases for $\mu B_{0}>0.5$.

Calculating the Hall response for a weaker magnetic field ($\mu B_{0}=0.1$)
we find qualitatively identical results. In Fig.~\ref{Fig9} we see two
maxima along $\theta =135^{\text{o}}$, the only difference being that the
area of increased Hall response is somewhat smaller in comparison with the
previous results for $\mu B_{0}=0.2$. For much stronger magnetic field ($\mu
B_{0}=1.0$) we find that there are still two maxima for $\theta =135^{\text{o%
}}$ with an increased Hall response. It is also clear from Fig.~\ref{Fig10}
that the difference in response within the junction is much larger for such
large magnetic fields.

Finally, we investigated the effect of circular corners in the Hall
junction, as this is also present in every experimental setup due to the
limited resolution of lithographic techniques with which these devices are
fabricated. To study this influence we considered circular corners with
different radii $a/W=0.0,0.1,0.2,0.3$ (see the inset of Fig.~\ref{Fig11}).
In Fig.~\ref{Fig11} the Hall response is given along the center of the
voltage probes, and it is clear that circular corners in the Hall bar
decrease the Hall response considerably. The response function $\widetilde{F}%
_{H}$ decreases with increasing radius $a/W$ near the center of the Hall
cross and $\widetilde{F}_{H}$ stays flat over a larger region in the center.
This is a consequence of the increased effective area of the Hall cross. For
small $a/W$ we found that in the center of the Hall cross the dependence on
the smoothness of the corners can be approximated by $\widetilde{F}%
_{H}\left( a\right) =\widetilde{F}_{H}\left( 0\right) $. Note also that $%
\widetilde{F}_{H}=F_{H}/W^{2}$, where $W^{2}$ was the Hall junction area in
the case of square corners. For circular corners this surface area should be
replaced by $\widetilde{W}^{2}>W^{2}$, where $\widetilde{W}$ is an
increasing function of the radius $a/W$. Scaling the response function
accordingly, i.e. $F_{H}/\widetilde{W}^{2}=\widetilde{F}_{H}\cdot \left(
W^{2}/\widetilde{W}^{2}\right) $, would lower the response function even
more. But we can conclude from this study that for small radius, i.e. $%
a/W=0.1$ , which is often a reasonable approximation for actual Hall
magnetosensors, the effect of circular corners on the response function
remains small, i.e. $<3\%$. The contourplot of the Hall response for this
geometry is given in Fig.~\ref{Fig12}, for the case of $a/W=0.2$, where the
resemblance with the square geometry in Fig.~\ref{Fig5} is apparent.
Consequently, the same analytical expression as for the case of square
corners, i.e. Eq.~(\ref{eq13}) can be used. The corresponding parameters are
given in Fig.~\ref{Fig7} by the open symbols. Notice that only for the
parameter $C_{4}$ major deviations (with respect to the square corner
result) are found for $\theta \thickapprox 45^{\text{o}}$ and $\theta
\thickapprox 135^{\text{o}}$ where the sharp dips are smoothed out.

\section{Asymmetric Hall cross}

The above results indicated that the sensitivity of a Hall bar is not
constant throughout the Hall cross. Thus, one can ask oneself whether it is
possible to enhance the sensitivity in certain parts of the Hall cross by
using a special cross geometry. Therefore, we investigated the influence of
asymmetry of the Hall cross, i.e. when the probes do not have the same
width, on the Hall response.

\subsection{Narrow voltage probes}

When the geometry of the Hall cross is such that the current probes with
width $W_{C}$\ are wider than the voltage probes with width $W_{V}$ ($%
W_{V}<W=W_{C}$)\ we obtained a response function which can be substantially
larger than $\widetilde{F}_{H}>0.5$, indicating that a more sensitive region
is created by narrowing the voltage probes. The Hall response along the
center of the voltage probes is given in Fig.~\ref{Fig13} for different
values of the voltage probe width: $W_{V}/W_{C}=0.3,0.5,0.7,1.0$. We took $%
\mu B_{0}=0.2$ and $r_{0}^{\prime }=0.1$. From this figure it is clear that
a more pronounced peak structure arises as the voltage probe width is
decreased. The contourplot for $\widetilde{F}_{H}$ of the geometry with $%
W_{V}/W_{C}=0.7$ is given in Fig.~\ref{Fig14}. Notice that the response
function was scaled as $\widetilde{F}_{H}=F_{H}/W^{2}$ with $W=W_{C}$. If we
scale the response function with the effective Hall junction area $%
W_{C}W_{V}<W^{2}$ this would lead to an even higher response function. From
this we may conclude that it is not necessary to narrow all the probes, but
that it is sufficient to do this only for the voltage probes in order to
enhance the sensitivity of the device.

\subsection{Asymmetric voltage probes}

We also investigated the situation when only one voltage probe is narrow ($%
W_{V}^{\prime }<W=W_{V}=W_{C}$) and all other voltage and current probes
have the same width. The result is a single-peak function along the voltage
probes, given by the curves in Fig.~\ref{Fig15}, where a more sensitive area
is created in the Hall junction close to the narrow probe. For comparison,
we also included the results for the other geometries. Compared to the
previous situation this geometry leads to a somewhat lower value for the
Hall response, but has considerable experimental advantage as only one of
the probes would have to be narrowed to obtain a sensitive region. The
contourplot of the Hall response for this configuration is given in Fig.~\ref
{Fig16}.

\subsection{Narrow current probes}

Finally, also the current probes were narrowed with respect to the voltage
probes ($W_{C}<W=W_{V}$). As we can see in Fig.~\ref{Fig17} the Hall
response increases with decreasing current probe width leading to a response
function which can be much larger than $\widetilde{F}_{H}>0.5$, and
indicating increased sensitivity of the sensor. In Fig.~\ref{Fig18} the
contourplot of the Hall response is plotted for this geometry with $%
W_{C}/W_{V}=0.7$ and we see that, analogous to the case where the voltage
probes were narrowed, there are two maxima which are shifted towards the
center of the narrowed probe.

\section{Conclusion}

In contrast to the ballistic regime (Ref.~\cite{ballistic}) where the Hall
response function is a step function $\widetilde{F}_{H}(x,y)=\frac{1}{W^{2}}%
\theta \left( W/2-\left| x\right| \right) \theta \left( W/2-\left| y\right|
\right) $, in the diffusive regime it is a smooth function which is constant
only near the center of the Hall cross and a decaying function which is
asymmetric and different from zero, although small, in the voltage and
current probes. From our study it is clear that different regions in the
Hall bar are more, or less, sensitive to the presence of a magnetic field.
In order to quantify this observation we determined the response function of
the Hall device. Lithographic fabrication techniques have a limited
resolution which led us to investigate the effect of circular corners in the
Hall junction. It was then observed that circular corners decrease the Hall
response significantly so that considerable attention should be paid to the
resolution with which these devices are fabricated. Bearing in mind the
fabrication of Hall sensors, several other geometric influences were also
investigated, and led to the conclusion that more sensitive areas can be
created in the Hall junction simply by narrowing the voltage or current
leads. Further experimental simplification can be obtained by narrowing only
one of the voltage probes as it appeared that this already creates a more
sensitive region in the junction.

{\bf Acknowledgements }Part of this work was supported by the Flemish
Science Foundation (FWO-Vl), the Inter-University MicroElectronics Center
(IMEC, vzw), the Concerted Action Programme (GOA) and the Inter-University
Attractions Poles (IUAP) research programme. Discussions with V. Schweigert
and A. Matulis are gratefully acknowledged.

\begin{figure}[tbp]
\caption{The Hall cross geometry. The current is injected along the $y$
direction.}
\label{Fig1}
\end{figure}

\begin{figure}[tbp]
\caption{The Hall sensor scans the surface (grey area), i.e. it moves across
the surface to determine the magnetic field distribution.}
\label{Fig2}
\end{figure}

\begin{figure}[tbp]
\caption{Potential profile arising due to the presence of a magnetic dot in
the center of the junction (indicated by the shaded circle) with strength $%
\protect\mu B_{0}=5.0$ and radius $r_{0}^{\prime }=r_{0}/W=0.3$. The values
on the contourplot are in units of $V_{0}$, the applied voltage. The current
is along the $y$ direction.}
\label{Fig3}
\end{figure}

\begin{figure}[tbp]
\caption{The Hall response for a magnetic dot ($\protect\mu B_{0}=0.2$ and $%
r_{0}^{\prime }=0.1$) displaced along the center of the current ($\protect%
\theta =90^{\text{o}}$) and voltage ($\protect\theta =0^{\text{o}}$) probes.
The curves represent Eq. (\ref{eq12}) for the corresponding angles. }
\label{Fig4}
\end{figure}

\begin{figure}[tbp]
\caption{Contourplot of the Hall response $\widetilde{F}_{H}$ in a Hall bar.
The scanning tip consists of a magnetic dot with strength $\protect\mu %
B_{0}=0.2$ and radius $r_{0}^{\prime }=0.1$. }
\label{Fig5}
\end{figure}

\begin{figure}[tbp]
\caption{The Hall response for a magnetic dot ($\protect\mu B_{0}=0.2$ and $%
r_{0}^{\prime }=0.1$) displaced along the axis at an angle $\protect\theta %
=0^{\text{o}},45^{\text{o}},$ $90^{\text{o}}$ and $135^{\text{o}}$ with the $%
x$ axis.}
\label{Fig6}
\end{figure}

\begin{figure}[tbp]
\caption{The values of the parameters in Eq. (\ref{eq13}) as a function of $%
\protect\theta $ for a Hall cross geometry with square corners (solid
symbols) and circular corners with $a/W=0.2$ (open symbols).}
\label{Fig7}
\end{figure}

\begin{figure}[tbp]
\caption{Dependence of the Hall response on the strength $\protect\mu B_{0}$
and the radius $r_{0}^{\prime }=r_{0}/W$ of the magnetic dot. The curves
shown are for $r_{0}^{\prime }=0.05,0.10,0.15,0.2,0.3,0.4,0.5$. The inset
shows the dependence of the parameters $A_{1}$ and $A_{2}$ (Eq. (\ref{eq14}%
)) on the radius of the dot. For small radii ($r_{0}^{\prime }<0.3$)\ there
exists a linear relation as shown schematically by the dotted and dashed
lines, respectively.}
\label{Fig8}
\end{figure}

\begin{figure}[tbp]
\caption{Position dependence of the Hall response $\widetilde{F}_{H}$ in a
Hall bar, as resulting from scanning a magnetic dot with strength $\protect%
\mu B_{0}=0.1$ and radius $r_{0}^{\prime }=0.1$.}
\label{Fig9}
\end{figure}

\begin{figure}[tbp]
\caption{Contourplot of the Hall response $\widetilde{F}_{H}$ in a Hall bar,
due to the presence of a magnetic dot with strength $\protect\mu B_{0}=1.0$
and radius $r_{0}^{\prime }=0.1$.}
\label{Fig10}
\end{figure}

\begin{figure}[tbp]
\caption{Circular corners in the Hall junction decrease the Hall response.
The response function is shown here as resulting from a magnetic dot ($%
\protect\mu B_{0}=0.2$ and $r_{0}^{\prime }=0.1$) which is displaced along
the center of the voltage probes for different values of $%
a/W=0.0,0.1,0.2,0.3 $.}
\label{Fig11}
\end{figure}

\begin{figure}[tbp]
\caption{Contourplot of the Hall response $\widetilde{F}_{H}$ in a Hall bar
with circular corners ($a/W=0.2$). The scanning tip consists of of a
magnetic dot with strength $\protect\mu B_{0}=0.2$ and radius $r_{0}^{\prime
}=0.1$.}
\label{Fig12}
\end{figure}

\begin{figure}[tbp]
\caption{A Hall cross geometry with narrow voltage probes, with respect to
the current probes ($W_{V}<W_{C}$), creates two much more sensitive regions
in the Hall junction, which can be seen from this two-peak function along
the center of the voltage probes for $W_{V}/W_{C}=0.3,0.5,0.7,1.0$. The Hall
response is scaled as follows $\widetilde{F}_{H}=F_{H}/W^{2}$ with $W=W_{C}$%
. }
\label{Fig13}
\end{figure}

\begin{figure}[tbp]
\caption{Contourplot of the Hall response $\widetilde{F}_{H}$ in a Hall bar
with narrow voltage probes ($W_{V}/W_{C}=0.7$). The scanning tip is a
magnetic dot with strength $\protect\mu B_{0}=0.2$ and radius $r_{0}^{\prime
}=0.1$.}
\label{Fig14}
\end{figure}

\begin{figure}[tbp]
\caption{One voltage probe is narrowed with respect to the current probes ($%
W_{V}^{\prime }<W_{V}=W_{C}$). This results in a single-peak function along
the center of the voltage probes (dotted curve). The dashed curve represents
the Hall response $\widetilde{F}_{H}=F_{H}/W^{2}$ (with $W=W_{C}$) where
both the voltage probes were narrowed $W_{V}^{\prime }=W_{V}<W_{C}$, while
the solid curve represents the symmetric case $W_{V}^{\prime }=W_{V}=W_{C}$.}
\label{Fig15}
\end{figure}

\begin{figure}[tbp]
\caption{Contourplot of the Hall response $\widetilde{F}_{H}$ in a Hall bar
with one narrow voltage probe $W_{V}^{\prime }<W_{V}=W_{C}$ ($W_{V}^{\prime
}/W_{V}=0.5$). The scanning tip is a magnetic dot with strength $\protect\mu %
B_{0}=0.2$ and radius $r_{0}^{\prime }=0.1$.}
\label{Fig16}
\end{figure}

\begin{figure}[tbp]
\caption{A Hall cross geometry with narrow current probes with respect to
the voltage probes ($W_{C}<W_{V}$) shows increased sensitivity as the width
is decreased. Here, the Hall response $\widetilde{F}_{H}=F_{H}/W^{2}$ (with $%
W=W_{V}$) along the center of the voltage probes is given for $%
W_{C}/W_{V}=0.3,0.5,0.7,1.0$.}
\label{Fig17}
\end{figure}

\begin{figure}[tbp]
\caption{Contourplot of the Hall response $\widetilde{F}_{H}$ in a Hall bar
with narrow current probes ($W_{C}/W_{V}=0.7$). The scanning tip is a
magnetic dot with strength $\protect\mu B_{0}=0.2$ and radius $r_{0}^{\prime
}=0.1$.}
\label{Fig18}
\end{figure}

\end{document}